\documentclass[aps,pre,preprint,groupedaddress,showpacs,floatfix]{revtex4}
\usepackage{amsmath}
\usepackage{amssymb}
\usepackage{graphicx}

\begin{document}

\title{Neuromorphometric characterization with shape functionals}
\author{Marconi Soares Barbosa}
\email[]{marconi@if.sc.usp.br}
\author{Luciano da Fontoura Costa}
\email[]{luciano@if.sc.usp.br}
\affiliation{Cybernetic Vision Research Group, GII-IFSC. Universidade de S\~ ao Paulo, S\~{a}o Carlos, SP, Caixa Postal 369, 13560-970, Brasil}
\author{Esmerindo de Sousa Bernardes}
\email[]{sousa@if.sc.usp.br}
\affiliation{Departamento de F\'isica e Ci\^encia dos Materiais. Universidade de S\~ ao Paulo, S\~{a}o Carlos, SP, Caixa Postal 369, 13560-970, Brasil}

\date{\today}

\begin{abstract}   
This work presents a procedure to extract morphological information from
neuronal cells based on the variation of shape 
functionals as the cell geometry undergoes a dilation through a wide interval of spatial
scales. The targeted shapes are alpha and beta cat retinal ganglion cells which
are characterized by different ranges of dendritic field diameter. Image functionals are expected to act as descriptors
of the shape, gathering relevant geometric and topological features of the complex cell form. We
present a comparative study of classification performance of additive  shape
descriptors, namely Minkowski functionals, and the non-additive multi scale
fractal. We found that the new measures perform efficiently the task of
identifying the two main classes, $\alpha$ and $\beta$, based solely on
scale invariant information, while also providing intraclass
morphological assessment. 
\end{abstract}


\maketitle

\section{introduction} 
Many natural phenomena are defined or influenced by the geometrical
properties of the involved elements, and vice-versa.  Examples of
such a shape-function relationship include the chemical properties of
proteins, the aerodynamic efficiency of wings, and the oxygen exchanges
through elaborated bronchic structures.
The close relationship between geometry and function provides strong motivation
for the geometrical analysis of natural objects. 
Yet, the state-of-the-art of geometrical characterization, an area
sometimes called morphometry or morphology, remains in a relatively incipient
stage where several competing, and often divergent, approaches coexist.  
While powerful methods have been used in physics to express
relevant geometrical properties, often including differential measurements
such as curvature, the situation remains particularly challenging in biology.

The relationship between neuronal shape and function has attracted increasing
attention due to its far reaching implications for basic neuroscience and for
medical applications. Neuron morphology has the special characteristic, that
it evolves during the developmental stage of the cell, being influenced by its
molecular environment and the history of synaptic activity~\cite{Ghosh:2002}.
The mature neuronal shape, together with its membrane electrical properties,
determine the electric conductance of the cell~\cite{Koch:2000} and account for
part of its electrophysiological characteristics such as firing patterns and
computational abilities~\cite{PEL02,FUK84,WAS81,KRAPP98}. Software packages, such as NEURON,
are available for modelling neuronal activity with basis on cable
theory~\cite{Rall:1977} which can be useful for analysis of real or virtual
neurons~\cite{Costa:1997,ASC01,Ascoli:2002}. At the same time, neuronal shape
can vary for different tissues, depending on a number of extracellular
factors~\cite{Kandel:1985,Ghosh:2002}. Ultimately, it determines the patterns of
connectivity and, consequently, the overall network computational
abilities. On the application side, tasks such as automated morphological
characterization of neuronal shape and the diagnosis of abnormalities, deserve
further investigation.  
                      
While the use of morphological tools was severely constrained until recently
by the cost of the relatively sophisticated systems needed to process images
and geometry, the continuing advances in computer software and hardware have
paved the way for an ever widening range of possible applications.
Consequently, more effective morphological concepts and methods have been
developed and reported in the literature, including the use of differential
geometry concepts such as multi-scale curvature and bending
energy~\cite{LU010,LU020,LU021}, methods from mathematical morphology such as
skeletonization~\cite{LU078,LU099}, as well as the recently reported framework
known as Integral-Geometry Morphological Image Analysis (MIA)
\cite{Raedt01,Raedt02,Mecke}. The latter approach involves the use of additive
shape functionals, i.e.  mappings that take shapes to single scalar values, in
terms of a parameter usually related to the spatial scale or time.  As far as
neuroscience is concerned, the contour of a neuronal cell has been shown to
possess a fractal structure~\cite{MOR89} and its multiscale fractal dimension
been used to characterize different morphological classes of neuronal
cells~\cite{LU101}.

Primarily motivated by the possibility of applying the methodology
proposed in \cite{Raedt01,Raedt02} as a novel and potentially useful tool for addressing
the problem of neuronal shape characterization and
classification, the present work also provides an assessment of those measures
considering a database of real biological data, namely camera lucida images of
cat ganglion neuronal cells.  In order to provide a comparative reference, the
multi-scale fractal dimension \cite{LU101}, itself a shape functional, is also
considered as a measure for shape characterization.  


Integral geometry provides an adequate mathematical framework for
morphological image analysis, having a core of useful theorems and formulae
that in some cases leads to analytical results for averages of image
functionals~\cite{Mecke} while also being quite efficient to implement
computationally~\cite{Raedt02}. Here, the class of functionals 
involved are restricted to additive, motion invariant and continuous,
called Minkowski functionals. These functionals are related  to usual
geometric quantities, for instance, in the Euclidean plane, to  area, perimeter and
connectivity or the Euler number, which expresses the number of holes
in a connected pattern such as the image of a neuronal cell.       

In order to describe geometrically one object, a set of measures (functionals)
is taken and the behaviour of these measures is monitored, as some 
control parameter is varied. In this work we compute additive
functionals in the plane as the contour of a neuronal cell image is
inflated by a parallel set dilation of radius $r$, the control parameter. The
non-additive multi scale fractal dimension is derived from one of those
computed additive functionals, at each radius of dilation, giving
important~\cite{LU083} complementary information.         

This article starts by presenting the adopted methodology, the considered
shape functionals, and the statistical procedure for cell identification.
The results, which are presented subsequently, clearly 
indicate that the new measures are efficient for distinguishing
morphologically the two functional classes alpha and beta as well as
revealing a strong morphological coherence in one of the classes.     

\section{Methodology}

\subsection{Additive shape functionals}
The morphological characterization in Euclidean plane by means of shape
functionals explores simple properties of convex sets. For these basic
geometric objects, such as triangle or an ellipse, we may evaluate a
change in area while the object undergoes a morphological dilation with 
the knowledge of its initial geometry. 
For example the change in area of a convex body $K$, after a parallel set dilation using a 2D ball
of radius $r$, can be expressed as   
\begin{equation}\label{eq_dila}
A(K_r)=A(K)+U(K)r+\pi r^2,
\end{equation}
where $A(K)$ and $U(K)$ stand for the initial area and perimeter of the object $K$
and $r$ is the dilation parameter. The process of taking parallel sets
generalises naturally to higher dimensions, while the change in hyper volume
preserves the general form \eqref{eq_dila} and is given by the Steiner
formula
\begin{equation}\label{eq_dilageral}
v^d(K_r)=\sum_{\nu=0}^d  \binom{d}{\nu} W_{\nu}^{(d)}(K)r^{\nu},
\end{equation} 
where the coefficients $W_{\nu}^{(d)}$ are referred to as {\it quermassintegrals} or
Minkowski functionals, \cite{Raedt01}. These functionals, as a generalisation of known
geometric quantities, are additive, motion invariant
and continuous. Moreover, a Theorem by Hadwiger~\cite{Raedt01,Mecke} states that these functionals form a
complete set of measures, with the above properties, on the set of convex bodies  
\begin{equation}
\phi(K)=\sum_{j=0}^{d}c_{j}W_{j}^{(d)}(K).
\end{equation}
Notwithstanding, the change of any, additive, motion invariant and continuous functional can be 
expressed, using a generalised Steiner formula~\cite{Mecke,Santalo}, in terms of the initial geometric information  
\begin{equation}\label{eq_general}
\phi(K_r)= \sum_{j=0}^{d}\sum_{k=0}^{d-j}c_j \binom{d-j}{k} W_{k+j}^{(d)}(K)r^k.
\end{equation}

The notion of connectivity number or Euler characteristic $\chi$ is central in
establishing the aforementioned properties of Minkowski functionals. The usual
definition of the connectivity from algebraic topology in two dimensions is
the difference between the number of connected $n_c $ components and the number 
of holes $n_h$, 
\begin{equation}
\chi(K)=n_c-n_h,
\end{equation}
while in three dimensions distinction should be made between two kind of
holes namely cavities $n_{hc}$ and handles(tunnels) $n_{hh}$  
\begin{equation}
\chi(K)=n_c-n_{hh}+n_{hc}.
\end{equation}
Integral Geometry provides an equivalent definition for
connectivity number of a convex set $K$ which is given by
\begin{equation}
\chi(K)=
      \begin{cases}
      1 & K \neq \varnothing \\
      0 & K = \varnothing.
      \end{cases}  
\end{equation}
Of great importance is its property of additivity 
\begin{align}\label{eq_additivity}
\chi(A)=&\chi(\cup_{i=1}^{l}K_i )=\sum_i\chi(K_i)-\sum_{i<j}\chi(K_i\cap K_j) 
+ \ldots \notag\\
&\ldots+(-1)^{l+1 }\chi(K_1\cap\ldots\cap K_l). 
\end{align}
Additivity and motion invariance is inherited by the
Minkowski functionals as they are related to the connectivity number by the
formulae
\begin{align}
&W_{\nu}^{(d)}(A)=\int_{\mathcal{G}}\chi(A\cap  E_{\nu})d\mu{E_{\nu}} \quad \nu=0,\ldots, d-1
\notag\\
&W_{d}^{(d)}(A)=\omega_d\chi(A),\quad \omega_d=\pi^{d/2}/\Gamma(1+d/2).
\end{align}
In the above expression $E_{\nu}$ stands for an $\nu$-dimensional plane in
$\mathrm{R}^{d}$. The integral is to be taken for all positions, induced by
isometries $\mathcal{G}$, of $E_{\nu}$ weighted by $d\mu(E_{\nu})$, the
kinematical density which is in turn related to the Haar measure on the
group of motions $\mathcal{G}$, see \cite{Raedt01,Mecke,Santalo}. 

To sum up, the Minkowski functionals $W_{\nu}^{(d)}(A)$, as a
generalisation of the usual procedure for volume determination, counts the
number of possible intersection of a $\nu$-dimensional plane with the domain
$A$. 

If one is to take advantage of the above additivity property all intersection
in \eqref{eq_additivity} must be taken into account. When working on a lattice,
there is a more expedient route, exploring the discrete nature of the images
and the additivity of the Minkowski functionals, which consists of a
decomposition of the 2D body $A$ into a disjoint collection of interior bodies,
open edges and vertices. Following the usual nomenclature we
denote the interior of a set $A$ by $\breve{A}=A/\partial{A}$. For an open
interior of a $n$-dimensional body embedded in a $d$-dimensional Euclidean space
there is the following expression for the Minkowski functionals, \cite{Raedt01}, 
\begin{equation}
W_{\nu}^{(d)}(\breve A)=(-1)^{d+n+\nu}W_{\nu}^{(d)}(A), \quad \nu=0,\ldots,d.
\end{equation}
We may then apply additivity and the lack of connectivity of open sets on the
lattice to determine the functionals for the body as a whole
\begin{equation}\label{eq_whole}
W_{\nu}^{(d)}(\mathcal{P})=\sum_{m} W_{\nu}^{(d)}(\breve{N}_m)n_m(\mathcal{P}), \quad \nu=0,\ldots,d.
\end{equation}  
Where $n_m(\mathcal{P})$ stands for the number of building elements of each type
$m$ occurring in the pattern $\mathcal{P}$.  
For a two dimensional space which is our interest for the present neuron
images,  we display in Table~\ref{tb_buildblocks} the value of Minkowski 
functionals for the building elements on a square lattice of pixels and
their direct relation to familiar geometric quantities on the plane. Using the
information presented in Table~\ref{tb_buildblocks}
and equation \eqref{eq_whole} we have
\begin{equation}\label{eq_func}
\quad A(\mathcal{P})=n_2, \quad U(\mathcal{P})=-4n_2+2n_1, \quad \chi(\mathcal{P})=n_2-n_1+n_0.
\end{equation}
So the procedure of calculating Minkowski functionals of a pattern $\mathcal{P}$ has
been reduced to the proper counting of the number of elementary bodies of each
type that compose a pixel (squares, edges and vertices) involved in the make
up of $\mathcal{P}$.

\begin{table*}[tb]
\begin{ruledtabular}
\begin{tabular}{lllll}
$m$ & $\breve{N}_m$  & $W_0^{(2)}=A(\breve{N}_m)$ & $W_1^{(2)}=\frac{1}{2} U(\breve{N}_m)$ & $W_2^{(2)}=\pi\chi(\breve{N}_m)$ \\ \hline
0 &  $\breve{P}$   & 0           & 0           & $\pi$    \\ \hline
1 &  $\breve{L}$   & 0           & $a$         & $-\pi$   \\ \hline
2 &  $\breve{Q}$   & $a^2$       & $-2a$       & $\pi$    \\ 
\end{tabular}
\caption{Minkowski functionals of elementary open bodies which compose a
  pixel $K$.\label{tb_buildblocks}}
\end{ruledtabular}
\end{table*}
In section \ref{Results} we describe typical results for the  evaluation of
the above presented additive functionals using an actual neuron image. The procedure involves the
implementation of, first an algorithm for the proper parallel set dilation throughout
all permitted radius on the square lattice and, second, of  an algorithm for the
calculation of Minkowski functionals by counting disjoint building elements
based on the formulae \eqref{eq_func}.  An efficient routine for
undertaking the latter is described in detail in \cite{Raedt01,Raedt02}.   

\subsection{Multi scale fractal dimension}\label{MSF}
As an example of a related non-additive functional, we add to the previous
measures the multi scale fractal, an approach which has been applied successfully
to neuromorphometry~\cite{LU083}. 
The notion of multi-scale fractal dimension refers to a quantitative characterization
of complexity and the degree of self-similarity at distinct spatial
scales, see \cite{LU083,LU101}. Intuitively, the fractal dimension indicates how much the curve extends
itself throughout the space. As a consequence, more intricate curves will cover
the surround space more 
effectively and will display a higher fractal dimension. This quantity is
calculated via the derivative of the logarithm of the changing interior area as the
neuron cell image undergoes a dilation. As such it is immediately derivable from
the first additive functional (Area) in a pixel based approach as opposed to a
curvature approach, see \cite{LU101}.  

\subsection{Implementation}
We have conducted the evaluation of the functionals described above on a 800MHz
ordinary personal computer running Linux. Both the algorithm for exact
dilations on the square lattice, according to~\cite{TAD01} and the pixel
based algorithm for the estimation of the Minkowski functionals~\cite{Raedt01}
was implemented in SCILAB-2.6. It took approximately 40s to calculate the
functionals for each dilation radius.

\section{Results}\label{Results}

We start by describing a measure that is perhaps the simplest
one but which holds important information not only on its own, but through
relation with the multi scalar fractal dimension described in section~\ref{MSF}. In Figure~\ref{Area}
we show the typical monotonically increasing curve of the interior area of a neuron cell
as its contour is inflated by a parallel set procedure. To capture the gross
structure of this measure we calculate the area under the curve, ${\sf sum}^a$, and
in a size independent manner, the radius ${\sf R}^{a}_{\frac{1}{2}}$ at which the 
area below this curve reaches half of its value at the end of dilation. The
fine structure is given by its standard deviation, ${\sf std}^a$.

In Figure~\ref{Per} we show a typical curve of the perimeter of the evolving
frontier of the neuron cell as the contour is inflated by a parallel set
procedure. It is important to observe that this measure is affected by
errors introduced by the discrete nature of the lattice of
pixels and the low resolution of the original image, which becomes more acute
in beta cells because of their reduced size. Nonetheless, this effect tends to
be less important after the tenth radii or so.  
To capture the gross structure of this measure, we calculate the area under
the curve ${\sf sum}^p$ and the size independent radius ${\sf R}^{p}_{\frac{1}{2}}$
at which the area below this curve reaches half of its overall value. The
fine structure is given simply by the standard deviation of the data, ${\sf std}^p$.

\begin{figure}[pt]
\includegraphics[scale=.45]{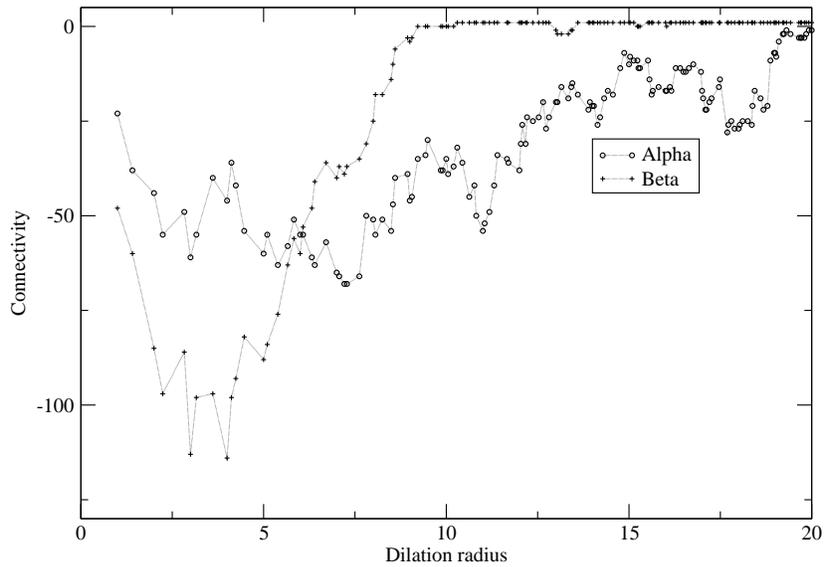}%
\caption{Comparison between the connectivity of typical alpha and beta
  cells showing a different range (in pixels) of complexity for the two class of neurons. \label{AB}}
\end{figure}

In contrast to the preceding measure the connectivity or the Euler
number of the neuron shape as it undergoes the dilation processes is
independent of the resolution of the image. It is a measure
restricted to the topology of the shape counting essentially the number of
holes at each radius of dilation no matter the holes are perfectly round or
not. The alterations of connectivity are subtle from step to step, which is  
reflected by the complex distribution of cusps in
Figure~\ref{Con}. The vast amplitude of scale for which there is an abrupt change
of connectivity is a measure of complexity  
of this type of cell. Figure~\ref{AB} shows the particular behaviour of the connectivity,
for samples from the two classes, as the cell shape undergoes a dilation. We take the gross information of this
measure by extracting the area under the interpolated curve, ${\sf sum_{interp}}$, and a
monotonicity index given by   
\begin{equation}
i_s=\frac{s}{s+d+p},
\end{equation}
where $s,d$ and $p$ counts respectively the number of times that the curve increase,
decrease or reach a plateau. This index characterizes a perfect
monotonically increasing curve when its value is 1 and reaching its minimum value
for a curve of high variability. This measure is designed specially to explore the multi-scale
nature of the neuronal complexity. The finer structure is captured solely by the
standard deviation ${\sf std_{diff}}$ of the difference between the
interpolated curve and the original data.     

\begin{figure}[pt]
 \includegraphics[scale=.4,angle=90]{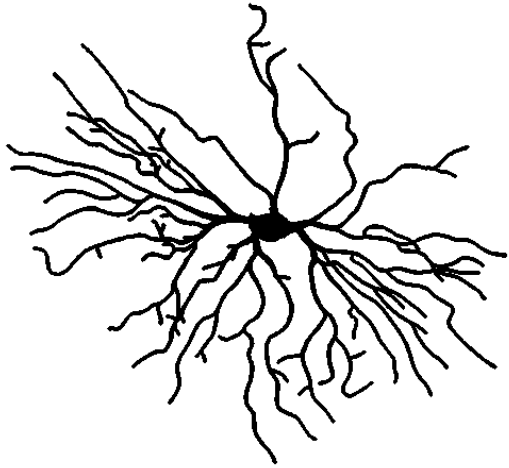}%
 \caption{Initial neuron image. Reprinted with permission from B.B. Boycott
   and H. W\"assle. J. Physiol(1974) 240 page 402. \label{BEG}}
\end{figure}

\begin{figure}[pt]
\includegraphics[scale=.4]{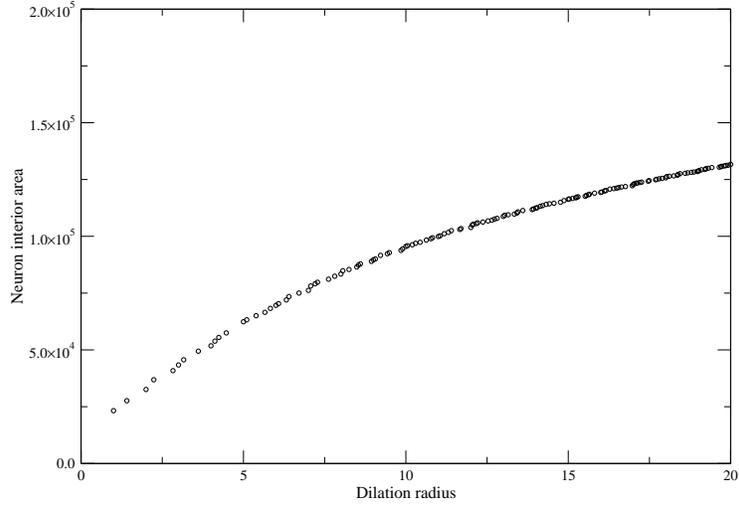}
\caption{Area (in pixels) as a function of the scale radius (in pixels) for a typical alpha
 neuronal cell given in Figure \ref{BEG}. Subtle information in this graphic is revealed only through
 the multi-scale fractal dimension. \label{Area}}
\end{figure}

\begin{figure}[pb]
\includegraphics[scale=.4]{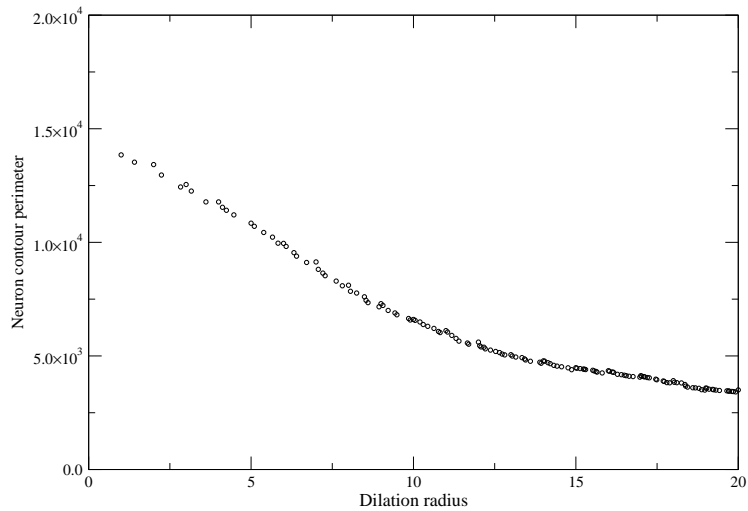}%
\caption{Perimeter (in pixels) as a function of the scale radius (in pixels) for the alpha
 neuronal cell of Figure \ref{BEG}. Note the expected initial decline and a
 visible fine structure associated with the disappearing and less frequently appearing holes. \label{Per}} 
\end{figure}

\begin{figure}[pt]
\includegraphics[scale=.4]{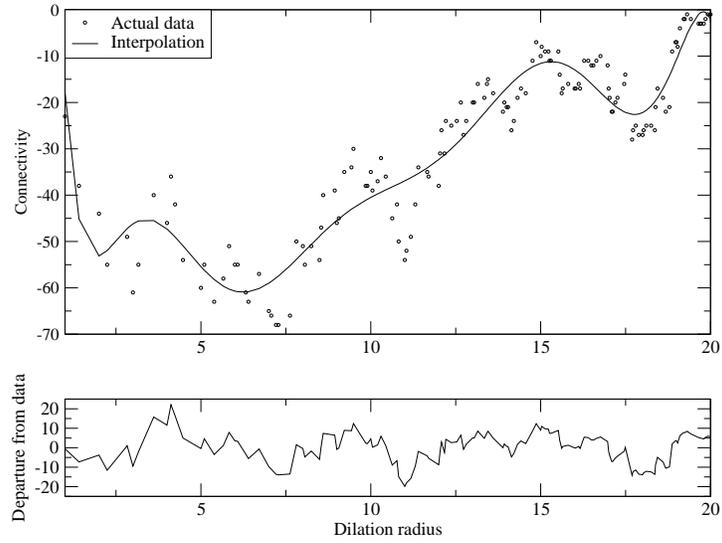}
\caption{Connectivity and the difference between data and interpolation for the
 typical alpha neuronal cell, Figure \ref{BEG}, as a function of parallel set
 dilation radius (in pixels). \label{Con}}
\end{figure}

\begin{figure}[pb]
\includegraphics[scale=.4]{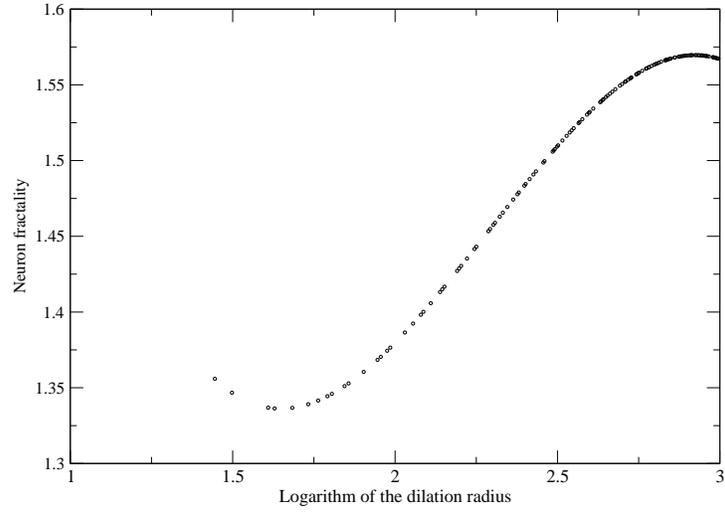}
\caption{Multi scale fractality of a neuronal type alpha cell. This attribute
  is obtained from the area functional and shows the fine structure that is
  not revealed by that functional alone. Dilation radius in
  pixels.~\label{Frac}}
\end{figure}

As a measure of complexity, the multi-scale fractal dimension has been experimentally found to be related to the
connectivity of a shape. Although this relationship is not straightforward, there might be a
correlation between these measures as commented bellow. For this image functional, we evaluate the maximum fractal
dimension, the mean fractality and the standard deviation, respectively {\sf max}, {\sf mean} and ${\sf std}^f$.   
A typical curve for the multi scale dimension is shown in Figure
\ref{Frac} for the same neuron appearing in Figure \ref{BEG}.

\begin{figure}[pt]
\includegraphics[scale=.4,angle=-90]{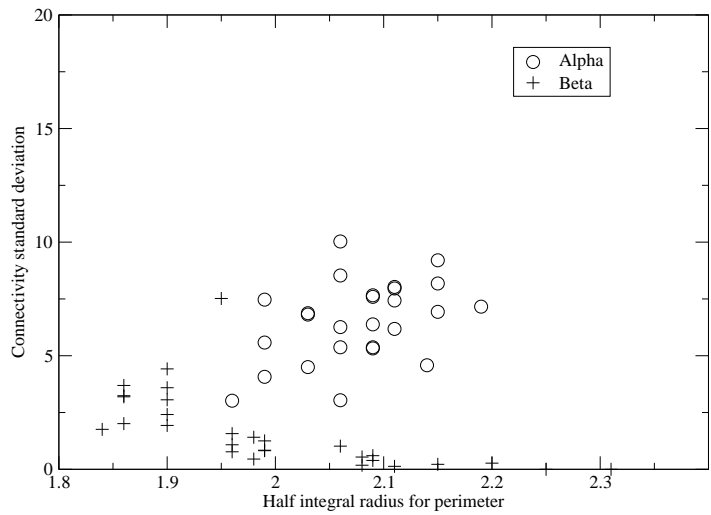}%
\caption{Clustering of alpha and beta cells based on $R_{1/2}^{\sf p}$ (in
  pixels) and connectivity measures.\label{Sep1}}
\end{figure}
\begin{figure}[pb]
\includegraphics[scale=.4,angle=-90]{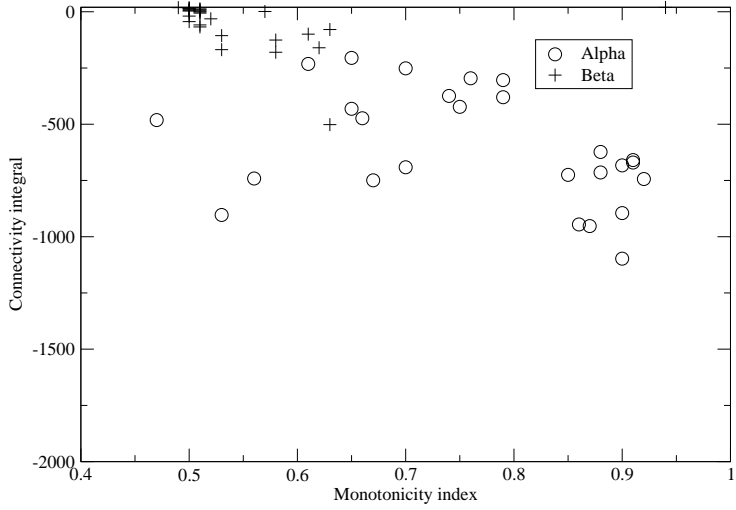}%
\caption{The feature space based on the index of monotonicity and 
 the integral of the connectivity (interpolated) curve, for both alpha and
 beta cells showing a strong clustering of the beta neuron class and a much
 dispersed alpha class. \label{Sep2}} 
\end{figure}

Among all the considered measurements, we found a good separation of
alpha and beta type cells for the feature space defined by the perimeter
half integral radius, ${\sf R}^p_{\frac{1}{2}}$, 
and the standard deviation for the fine structure of connectivity, ${\sf std_{diff}}$.   
Figure~\ref{Sep1} shows the obtained alpha and beta clustering with both
class exhibiting similar dispersion. Another good result for morphological
characterization was obtained for the connectivity or the Euler number of the cell shape. A feature space
involving the index of monotonicity and the integral of the interpolated
connectivity curve is presented in Figure~\ref{Sep2}. More efficient than the
former in separating the two classes, these measures produce a well localised 
clustering of beta cells characterising its geometrical intricacies. This
result suggests that alpha class is 
indeed a more homogeneous category while the beta class may have a morphological subclass structure.

\begin{figure}[pb]
\includegraphics[scale=.4,angle=-90]{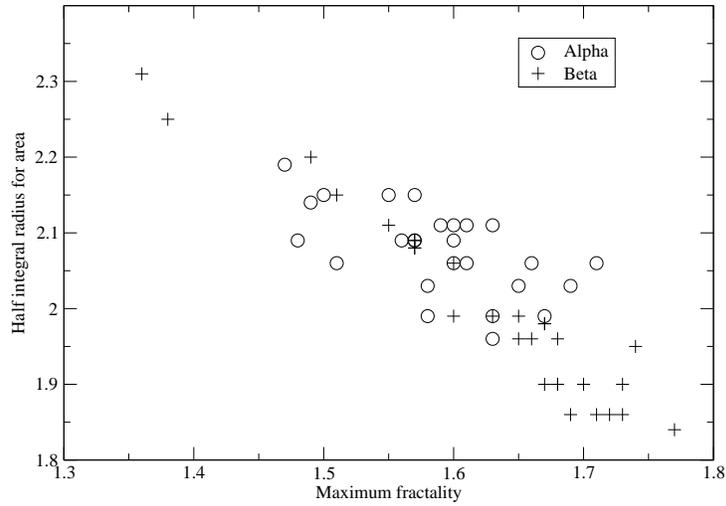}%
\caption{A not decisive feature space: related measures with high
 (anti-)correlation. Half integral radius in pixels.\label{Sep3}}
\end{figure}

Table \ref{corr} shows the correlation coefficients for the
twelve measurements considered in this work. Of special interest is the
correlation between fractality and connectivity, measures that have experimentally
been found to represent complementary but not redundant measures of
complexity. Figure~\ref{Sep3} shows a combination of two measures to form a
feature space which in this case shows a poor separation of classes which is
in accordance with the high anti-correlation of the involved measures
presented in Table~\ref{corr}. Unusually high correlation appears between some measures notably as occurring
between area and perimeter, suggesting a specific tendency that seems to be particular to the
type of data (neurons) and not a general rule.  

\section{Conclusions}   
The use of additive shape functionals has been recently considered for the
characterization of the geometrical properties of several physical
objects~\cite{Raedt01, Mecke}. The current article explored the use of a representative
set of such functionals, namely the area, perimeter and connectivity, for
the characterization  of neural shapes represented in terms of a whole set
of parallel expansions.  The multi scale fractal dimension, a non-additive
shape functional, was also considered as a standard for comparison.  


All the adopted shape functionals consist of functions of the dilating
radius of each parallel body. For the sake of efficience, the following
compact subset of global features was selected: the area
under the functionals, the value of radius where the area reaches its
half-value, the standard deviation of the functionals, as well as a novel
measurements expressing the monotonicity of the functionals and the
decomposition of the connectivity functional in terms of a low and high
variation signals.

All two-by-two combinations of these measures were investigated
visually in order to identify the combinations of features leading to more
pronounced separations between the two classes of considered neural cells,
namely cat retinal ganglion cells of type alpha and beta. The coefficients
of correlation of each pair of measures were also estimated and analysed,
indicating decorrelation between several of the considered features. The
obtained results confirmed a differentiated potential of each measurement 
for neural cell clustering, with the features derived from the
connectivity functional accounting for the best separation between
classes. However, further investigations and comparison to more established
and recent methods (such as in \cite{Mizrahi:2000}) based explicitly on dendritic morphology
will be necessary to reveal the best realization of the novel methodology
proposed in this article. The biological implications of such results are that the two 
type of cells differ in the distribution of holes (defined by the
respective dendritic arborizations) for different spatial scales. The
obtained clusters indicated that the alpha cells exhibit less uniform
geometrical properties than the beta, suggesting the existence of
morphological subclasses.

\begin{acknowledgments}
The authors are grateful to FAPESP (processes 02/02504-01,
99/12765-2 and 96/05497-3) and to CNPQ (process 301422/92-3) for
financial support.   
\end{acknowledgments}

\begin{figure}[pb]
\includegraphics[scale=.4,angle=90]{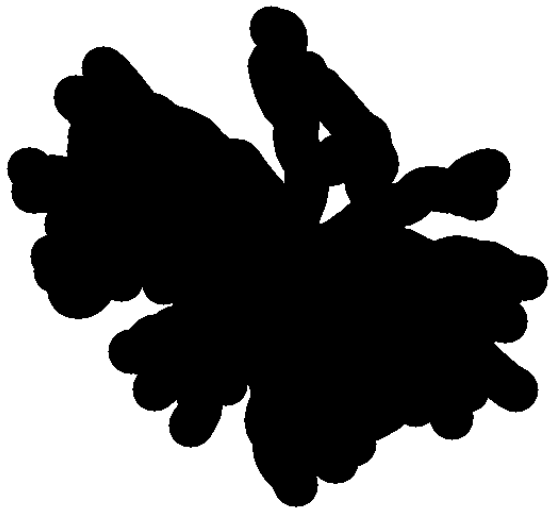}%
\caption{The same neuron image after the parallel set dilation. \label{END}} 
\end{figure} 

\begin{turnpage}
\begin{table*}[ht]
\begin{ruledtabular}
\begin{tabular}{l|rrr|rrr|rrr|rrr}
&\multicolumn{3}{c|}{\sf Area} & \multicolumn{3}{c|}{\sf Perimeter}& \multicolumn{3}{c|}{\sf Connectivity}&\multicolumn{3}{c}{\sf Fractality}\\\hline
 &${\sf std}^a$&${\sf sum}^a$&${\sf R}^a_{1/2}$&${\sf std}^p$&${\sf sum}^p$&${\sf R}^p_{1/2}$&${\sf std_{diff}}$&${\sf sum_{interp}}$&{\sf $i_s$}&${\sf std}^f$&{\sf max}&{\sf mean}\\\hline
${\sf std}^a$& 1& & & & & & & & & & & \\\hline 
${\sf sum}^a$& {\bf 0.96}& 1& & & & & & & & & & \\\hline 
${\sf R}^a_{1/2}$& {\bf 0.51}&       0.41& 1& & & & & & & & & \\\hline 
${\sf std}^p$& {\bf 0.86}& {\bf 0.96}&       0.33& 1& & & & & & & & \\\hline 
${\sf sum}^p$& {\bf 0.97}& {\bf 0.99}&       0.45& {\bf 0.94}& 1& & & & & & & \\\hline 
${\sf R}^p_{1/2}$&       0.13&       0.01& {\bf 0.78}&       -0.05&       0.03& 1& & & & & & \\\hline 
${\sf std_{diff}}$& {\bf 0.78}& {\bf 0.88}&       0.15& {\bf 0.90}& {\bf 0.87}&       -0.27& 1& & & & & \\\hline 
${\sf sum_{interp}}$& {\bf -0.79}& {\bf -0.92}&       -0.22& {\bf -0.98}& {\bf -0.90}&       0.17& {\bf -0.92}& 1& & & & \\\hline 
{\sf $i_s$}&       0.33&       0.44&       0.34& {\bf 0.52}&       0.44&       -0.02&       0.43& {\bf -0.57}& 1& & & \\\hline 
${\sf std}^f$&       -0.24&       -0.27&       0.16&       -0.19&       -0.27& {\bf 0.58}&       -0.40&       0.27&       -0.34& 1& & \\\hline 
{\sf max}&       -0.34&       -0.18& {\bf -0.86}&       -0.01&       -0.23& {\bf -0.65}&       0.05&       -0.07&       -0.22&       0.10& 1& \\\hline 
{\sf mean}&       0.14&       0.30& {\bf -0.63}&       0.39&       0.26& {\bf -0.88}& {\bf 0.55}&       -0.49&       0.22& {\bf -0.64}& {\bf 0.63}& 1\\\hline 
\end{tabular}
\caption{Correlation coefficients for the set of the measures extracted from
 Area, Perimeter, Connectivity and Fractality: Standard deviations (${\sf
 std}^{a,p,f}$,${\sf std_{diff}}$), Integrals (${\sf sum}^{a,p}$, ${\sf sum_{interp}}$),
 Half radii (${\sf R}^{a,p}_{1/2}$), the monotonicity index ({\sf $i_s$}),
 mean value ({\sf mean}) and max value ({\sf max}). Bold face for absolute values of correlation above 0.5. \label{corr}}
\end{ruledtabular}
\end{table*}
\end{turnpage}

\clearpage
 
\bibliography{gatorev}

\end{document}